# Handling an uncertain control group event risk in non-inferiority trials: non-inferiority frontiers and the power-stabilising transformation

**Running head:** Non-inferiority frontiers

**Word count:** 4133

**Authors:** Matteo Quartagno, A. Sarah Walker, Abdel G. Babiker, Rebecca M. Turner, Mahesh K.B. Parmar, Andrew Copas, Ian R. White

**Affiliations:** Institute for Clinical Trials and Methodology, University College London, 90 High Holborn, WC1V 6LJ

**Corresponding author:** Matteo Quartagno, MRC Clinical Trials Unit, University College London, 90 High Holborn, Second Floor. WC1V 6LJ, London, UK.
Email: m.quartagno@ucl.ac.uk

**Abstract:**

**Background.** Non-inferiority trials are increasingly used to evaluate new treatments expected to have secondary advantages over standard of care, but similar efficacy on the primary outcome. When designing a non-inferiority trial with a binary primary outcome, the choice of effect measure for the non-inferiority margin (e.g. risk ratio or risk difference) has an important effect on sample size calculations; furthermore, if the control event risk observed is markedly different from that assumed, the trial can quickly lose power or the results become difficult to interpret.

**Methods**. We propose a new way of designing non-inferiority trials to overcome the issues raised by unexpected control event risks. Our proposal involves specifying a "non-inferiority frontier", i.e. a curve defining the most appropriate non-inferiority margin for each possible value of control event risk. We propose a fixed arcsine difference frontier, using the power-stabilising transformation for binary outcomes. We propose and compare three ways of designing a trial using this frontier: testing and reporting on the arcsine scale; testing on the arcsine scale but reporting on the risk difference or risk ratio scale; and modifying the margin on the risk difference or risk ratio scale after observing the control event risk according to the power-stabilising frontier.

**Results.** Testing and reporting on the arcsine scale leads to results which are challenging to interpret clinically. For small values of control event risk, testing on the arcsine scale and reporting results on the risk difference scale produces confidence intervals at a higher level than the nominal one or non-inferiority margins that are slightly smaller than those back-calculated from the power-stabilising frontier alone. However, working on the arcsine scale generally requires a larger sample size compared to the risk difference scale. Therefore, working on the risk difference scale, modifying the margin after observing the control event risk, might be preferable, as it requires a smaller sample size. However, this approach tends to slightly inflate type I error rate; a solution is to use a lower significance level for testing, although this modestly reduces power.

When working on the risk ratio scale instead, the same approach based on the modification of the margin leads to power levels above the nominal one, maintaining type I error under control.

**Conclusions.** Our proposed methods of designing non-inferiority trials using power-stabilising non-inferiority frontiers make trial design more resilient to unexpected values of the control event risk, at the only cost of requiring larger sample sizes when the goal is to report results on the risk difference scale.

## 1. Introduction

Often a new treatment is expected not to have greater efficacy than the standard treatment, but to provide advantages in terms of costs, side-effects or acceptability. Here, a non-inferiority trial[1] can test whether the new treatment's efficacy is not unacceptably lower than standard treatment, and also where relevant guarantee that a minimum acceptable treatment effect relative to placebo is preserved, whilst providing sufficient evidence of superiority on secondary outcomes to support its use. Non-inferiority designs have been increasingly used in recent years[2].

A critical design choice is the non-inferiority margin, which is the largest acceptable loss of efficacy[3]. Considerations regarding margin choice depend on the type of primary outcome. We focus here on binary outcomes, for which either absolute[4] (risk difference) or relative[5] (risk ratio) margins can be defined. For example, the Food and Drug Administration guidelines[6] suggest that for licensing trials the results from placebo-controlled trials evaluating the standard treatment might inform margin choice, using the lower bound of the confidence interval for the estimated effect vs placebo, most often using the absolute scale. In other situations, the goal might be to preserve a certain proportion of the effect of the standard relative to placebo, which can be formulated as either an absolute or relative margin.

In both cases, the expected control arm (standard treatment) event risk plays a very important role in the choice of the non-inferiority margin[7]. However, at trial completion, the actual control event risk can differ considerably from the expected one. This can occur when prior information was not correct, for example when standard of care has improved over years[8] or because a slightly different sub-population was recruited[4] or because additional aspects of care (or a Hawthorne effect) influence outcomes in the control group. This can have serious consequences on the power, and hence the interpretation, of the trial, particularly when the expected control event risk is very large (e.g. >90%) or small (<10%): the latter is common in non-inferiority trials where existing treatments are often highly effective precluding demonstrating superiority of a new treatment on the primary endpoint.

For example, for control risk <0.5, the sample size needed to achieve 90% power under a 5% non-inferiority margin on the risk difference scale (one-sided alpha=2.5%) increases with the control event risk (Figure (a) in the additional material online); hence, if the control event risk is larger than anticipated, this reduces the power of the trial to demonstrate non-inferiority (Figure (b)). The opposite occurs when working on the risk ratio scale, so that a lower than expected control event risk reduces power. We discuss a specific example illustrating this below (the OVIVA trial[9]). Furthermore, higher power than designed may not actually aid interpretation. For example, Mauri and D'Agostino[10] discuss the ISAR-safe[11] non-inferiority trial, where the observed control event risk was much lower than originally expected. The results provided strong evidence of non-inferiority based on the pre-specified non-inferiority margin as a risk difference, but they were also consistent with a three-fold increase in risk, and so the authors did not conclude non-inferiority.

Here we propose a new method of designing non-inferiority trials, which protects against a lower or higher than expected control event risk, preserving power and interpretability of results.

## 2. The non-inferiority frontier

Assume we want to test whether a new treatment $T_1$ is non-inferior to the standard treatment $T_0$. The primary (binary) outcome is an unfavourable event, e.g. death or relapse within one year from randomisation. Let:

- $\pi_1, \pi_0$ be the true incidences in the experimental and control groups respectively

- $\pi_{e1}, \pi_{e0}$ be the expected incidences assumed in the sample size calculation. Usually $\pi_{e1} = \pi_{e0}$ but occasionally[4] studies are designed with $\pi_{e1} < \pi_{e0}$ or $\pi_{e1} > \pi_{e0}$.

- $\pi_{f1}$ be the largest acceptable incidence in the experimental group if the control group incidence is $\pi_{e0}$. In a trial with an unfavourable outcome, $\pi_{f1} > \pi_{e0}$.

- $\delta$ be the non-inferiority margin, defined as $\delta = \pi_{f1} - \pi_{e0}$ if the risk difference scale is used, and $\delta = \log(\pi_{f1}/\pi_{e0})$ if the (log-)risk ratio scale is used.

- $n_1, n_0$ be the sample sizes, with allocation ratio $r = n_1/n_0$.

Several recommendations have been given regarding choice of the most appropriate non-inferiority margin[3,6], involving both clinical and statistical considerations. Whilst sample size calculations allow for stochastic variation between the true control event risk $\pi_0$ and its final observed estimate $\hat{\pi}_0$, they do not allow for substantial misjudgment in the envisaged truth.

We therefore argue that it is insufficient to define non-inferiority in terms of a single margin $\delta$; it is instead preferable, at the design stage, to define a curve associating a specific margin $\delta_{\pi_0}$ to each possible value of control event risk $\pi_0$. We call this the *Non-Inferiority Frontier*.

## 2.1. Risk Difference vs. Risk Ratio

The standard design, assuming a single non-inferiority margin $\delta$ irrespective of $\pi_0$, corresponds to a fixed risk difference or fixed risk ratio frontier. These frontiers are shown in Figure 1. The region underneath the golden line is the non-inferiority region assuming a fixed risk difference frontier; whatever the control event risk, the new treatment is judged non-inferior if $\pi_1 - \pi_0 < 0.05$. Similarly, the region below the blue line is the non-inferiority region assuming a constant risk ratio frontier.

The choice of frontier is important even when the expected control event risk is correct, i.e. $\pi_{e0} = \pi_0$. As shown by Figure (a) and (b) in the additional material, power and sample size calculations using different analysis scales give very different answers even when the assumed $\pi_{f1}$ and $\pi_{e0}$ are the same.

## 2.2. The power-stabilising non-inferiority frontier

We propose a third choice of frontier, the fixed arcsine difference[12,13] frontier, i.e. constant $\text{asin}(\sqrt{\pi_{f1}}) - \text{asin}(\sqrt{\pi_{e0}})$. Although the arcsine difference is more difficult to interpret than other measures, its great advantage is that its asymptotic variance is independent of $\pi_0$. Hence, when using a fixed arcsine difference frontier, the sample size and power calculations are approximately unaffected by $\pi_{e0} - \pi_0$. We therefore call this the *Power-Stabilising Non-Inferiority Frontier*, represented by the dark green line in Figure 1.

### *2.3. Choosing the non-inferiority frontier*

The most appropriate non-inferiority frontier must be chosen using clinical, as well as statistical, arguments. If the investigators' only interest lies in the single binary efficacy outcome, an increase in event risk from 5% to 10% can be considered as undesirable as an increase from 45% to 50%; in both, the experimental treatment leads to 50 more events per 1000 patients and a fixed risk difference frontier might be appropriate. However, many investigators would feel that the former increase is more important than the latter. This could be justified by arguing that a relative effect measure is more likely to be transportable to other outcomes. In this case, as the control event risk increases, we might tolerate a larger absolute increase in intervention event risk. However, as shown in Figure (a), with the risk ratio frontier the maximum tolerable absolute difference quickly becomes very large if the control event risk was badly underestimated. The power-stabilising frontier is a good compromise. As an example, the OVIVA[9] trial aimed to determine whether oral antibiotics were non-inferior to intravenous antibiotics to cure bone and joint infections. Intravenous antibiotics were the standard based on historical precedent, not evidence. Based on pilot data from one tertiary referral centre, researchers expected a low control event risk of treatment failure ($\pi_{e0} = 5\%$); given this, they were happy to tolerate up to a 10% event risk for the experimental treatment, because of its substantial advantages (e.g reduced line complications, earlier hospital discharge), i.e. a 5% absolute margin. However, the observed pooled event risk across 29 centres of varying sizes was much higher ($\hat{\pi}_0 = 12.5\%$); assuming this reflected the control group risk, they were happy to tolerate an experimental event risk larger than implied by a fixed risk difference frontier ($\pi_{f1} = 17.5\%$). As the risk ratio increases with control risk, a fixed risk ratio frontier ($\pi_{f1} = 25\%$) was an alternative in this case. However, the investigators decided that the maximum tolerable experimental event risk given $\pi_0 = 12.5\%$ was $\pi_{f1} = 20\%$, which is close to the arcsine frontier ($\pi_{f1} = 19.5\%$).

Another aspect to consider, when choosing the frontier, is that sample size calculations give very different answers when working on different scales. In an example trial with one-sided α=2.5%, power=90%, $\pi_{e0} = 5\%$ and $\pi_{f1} = 10\%$, the sample size to show non-inferiority on the arcsine scale (568 patients/group) is larger than on the risk difference scale (400 patients/group; 5% absolute margin); hence, choosing the arcsine frontier may require up to 40% more patients. However, the sample size required to show non-inferiority on the risk ratio scale is larger still (832 patients/group; 2-fold relative risk margin).

## 3. Implementation

In a standard non-inferiority trial, results can be interpreted against fixed risk difference or risk ratio frontiers. There are several ways we could instead design a trial under the power-stabilising frontier. We introduce them here and provide an illustrative analysis example in Appendix B.

### 3.1. Test and report on the arcsine scale

The simplest solution is to design the trial pre-specifying the non-inferiority margin on the arcsine difference scale; it is then sufficient to test non-inferiority at this fixed margin and report a point estimate and confidence interval on the arcsine scale, regardless of the final observed control event risk. However, such results are not easily interpretable and are unlikely to be clinically acceptable.

### 3.2. Test on the arcsine scale, report on the risk difference scale

A second possibility is to perform the test on the arcsine scale, but report results on the risk difference (or risk ratio) scale. The problem here is that the test statistic may not correspond to the relationship of the margin to the confidence interval. We propose two ways to resolve this, and we present them for the risk difference scale, although they could be easily adapted to the risk ratio scale; given an estimated arcsine difference $\widehat{AS}$ with associated standard error $\hat{\sigma}_{AS}$, a fixed non-inferiority margin on the arcsine difference scale $\delta_{AS}$ and an estimated risk difference $\widehat{RD}$ with standard error $\hat{\sigma}_{RD}$:

*(i) Back calculation of margin.*

1) Calculate the Z statistic for the arcsine scale test:

$$Z_{AS} = \frac{\widehat{AS} - \delta_{AS}}{\hat{\sigma}_{AS}}$$

2) Calculate for what non-inferiority margin $\delta_{RD}$ we get the same Z statistic when testing on the risk difference scale:

$$\delta_{RD} = \widehat{RD} - Z_{AS} * \hat{\sigma}_{RD}$$

3) Report the confidence interval on the risk difference scale and p-value of the test for non-inferiority at margin $\delta_{RD}$:

$$p = \Phi^{-1}(Z_{AS}) \quad CI(1-\alpha) = (\widehat{RD} - z_{1-\alpha} * \hat{\sigma}_{RD} \,;\, \widehat{RD} + z_{1-\alpha} * \hat{\sigma}_{RD})$$

*(ii) Back calculation of significance level and modification of margin.*

1) Calculate the non-inferiority margin $\delta_{RD}^*$ on the risk difference scale corresponding to $\delta_{AS}$ on the arcsine scale for the observed value of control risk $\hat{\pi}_0$:

$$\delta_{RD}^* = sin(asin(\sqrt{\hat{\pi}_0}) + asin(\sqrt{\pi_{f1}}) - asin(\sqrt{\pi_{e0}}))^2 - \hat{\pi}_0$$

2) Calculate the Z statistic $Z_{RD}$ for the test on the risk difference scale:

$$Z_{RD} = \frac{\widehat{RD} - \delta_{RD}^*}{\hat{\sigma}_{RD}}$$

3) Calculate at what significance level $\alpha^*$ the test using $Z_{RD}$ would be equivalent to a α-level test using $Z_{AS}$:

$$z_{1-\alpha^*} = z_{1-\alpha} \frac{Z_{RD}}{Z_{AS}}$$

4) Report $(1 - \alpha^*)$ confidence interval on the risk difference scale and p-value of the test for non-inferiority at margin $\delta_{RD}^*$:

$$p = \Phi^{-1}(Z_{AS}) \quad CI(1 - \alpha^*) = (\widehat{RD} - z_{(1-\alpha^*)} * \hat{\sigma}_{RD} \; ; \; \widehat{RD} + z_{(1-\alpha^*)} * \hat{\sigma}_{RD})$$

Both approaches are potentially valid; when $\pi_0 < 50\%$, the adjustment is generally small and, most notably, confidence levels reported are larger than the nominal $(1 - \alpha)$. One difficulty with this approach is that sample size might be impractically large, particularly for small values of control event risk, when the goal is to report on the risk difference scale, for the reasons discussed in Section 2.3. Conversely, since sample size required to prove non-inferiority on the risk ratio scale is larger than on the arcsine scale, the non-inferiority margin $\delta_{RR}$ or the significance level $\alpha^*$ may be unacceptably large when the goal is to report on the risk ratio scale.

### *3.3. "Conditionally modify margin": modify non-inferiority margin after observing control group event risk*

Our favoured proposal is to design the trial using a standard risk difference or risk ratio margin $\delta$ and then modify the margin to $\delta^*$ only if the observed event risk $\hat{\pi}_0$ differs by more than a certain threshold $\epsilon$ from the expected $\pi_{e0}$. Specifically:

- At trial completion we observe $\hat{\pi}_0$;

- If $|\hat{\pi}_0 - \pi_{e0}| > \epsilon$ (risk difference scale) or $|\log(\hat{\pi}_0/\pi_{e0})| > \epsilon$ (risk ratio scale), then:

    - Find $\pi_{f1}^*$ that solves $\operatorname{asin}\left(\sqrt{\pi_{f1}^*}\right) - \operatorname{asin}(\sqrt{\hat{\pi}_0}) = \operatorname{asin}(\sqrt{\pi_{f1}}) - \operatorname{asin}(\sqrt{\pi_{e0}})$;

    - Modify non-inferiority margin to $\delta^* = \pi_{f1}^* - \hat{\pi}_0$ (risk difference) or $\delta^* = \log(\frac{\pi_{f1}^*}{\hat{\pi}_0})$ (risk ratio);

- o Test non-inferiority at margin $\delta^*$;
- Otherwise do not modify margin and test non-inferiority at $\delta$.

This approach, while preserving the simplicity in interpreting non-inferiority against risk differences or risk ratios, potentially helps preserve power and interpretability when the true control event risk is badly misjudged by modifying $\delta$ according to the power-stabilising frontier. Differently from the method in Section 3.2(ii), the margin is only modified when the observed control risk differs substantially from its expectation. However, since the margin is modified in a data-dependent way, the method is potentially prone to inflation of type I error. We explore this next.

### *3.4. Type I error and power of the "conditionally modify margin" method*

We simulate 100000 datasets for a range of designs and true incidences, starting from a base-case scenario and then investigating alternatives, changing simulation parameters one-by-one (Table 1), appropriately calculating sample size from the design parameters in Table 1 and the formulae in the additional material. Since sample size calculations give very different answers when using risk ratio or risk difference, we generate different datasets for the two effect measures.

*Type I Error.*

We consider 40 data-generating mechanisms for each scenario, with $\pi_0$ ranging between 0.5% and 20%, and $\pi_1$ derived under the non-inferiority null from the arcsine rule:

$$\text{asin}(\sqrt{\pi_1}) - \text{asin}(\sqrt{\pi_0}) = \text{asin}(\sqrt{\pi_{f1}}) - \text{asin}(\sqrt{\pi_{e0}}).$$

This is the appropriate data-generating mechanism for evaluating type I error assuming the power-stabilising frontier holds. We compare four different analysis methods:

1) Do not modify margin: simply test non-inferiority with margin $\delta$ on the risk difference/ratio scale;

2) Modify margin, with $\epsilon = 5\%$ for risk difference or $\log(2)$ for log risk ratio.

3) Modify margin, with $\epsilon = 2.5\%$ for risk difference or $\log(1.5)$ for log risk ratio.

4) Modify margin, with $\epsilon = 1.25\%$ for risk difference or $\log(1.25)$ for log risk ratio.

*Base-case.* Figure 2 shows the results of these simulations, designing and analysing the data on a risk difference (left) or risk ratio (right) scale. Given our chosen non-inferiority frontier, "do not modify margin" leads to inflated type I error rate if the control event risk is lower or higher than expected using the risk difference or risk ratio respectively. The three "conditionally modify margin" procedures are identical to "do not modify margin" in a small region around the expected control event risk; the width of this region is directly proportional to the magnitude of $\epsilon$. For $\pi_0 > 10\%$, the margin is almost always modified (Figure (c) in additional material), and the "conditionally modify margin" procedures have the same level of type I error. Using the risk ratio, this level is below the nominal 2.5%, while with the risk difference it is just above 3.5%.

Comparing the strategies with different $\epsilon$, the procedure using the smallest threshold seems preferable irrespective of the scale used. In particular, when using risk ratios, it leads to a type I error always below 2.5%, while with risk difference the rate remains slightly inflated at about 4% in some areas, particularly with low incidences.

*Other data-generating mechanisms.* Figure 3 shows the results for the alternative scenarios, using procedure 4 only, i.e. "conditionally modify margin" with the smallest threshold (other procedures in Figures (d)-(e) in additional online material). Allocation ratio (alternatives 5 and 6) has a greater impact than other factors, because with more patients allocated to control, the estimated risk is affected by less error. However, in general, conclusions are not altered substantially.

*Power.*

We again vary $\pi_0$ between 0.5% and 20%, but this time under the non-inferiority alternative with $\pi_1 = \pi_0$.

*Base-Case.* Under "do not modify margin", power is substantially reduced if $\pi_0$ is higher (risk difference) or lower (risk ratio) than expected (Figure 2). Using risk ratio, power of any of the "conditionally modify margin" methods is always either above the nominal 90% or above the power of the "do not modify margin" procedure. The only exception is with $\pi_0$ lower than expected when using risk difference; nevertheless, power remains close to 80% even in this scenario. Interestingly, the procedure with the smallest threshold is the only one not achieving the nominal power when the control event risk is correct, possibly because the margin is at times modified even when risk differs from the expected only because of random variation.

*Alternatives.* Figure 3 shows the results under the alternative scenarios using procedure 4. The greatest difference from the base-case configuration is in the scenario where the experimental treatment has higher efficacy than the control (Alternative 2), particularly for small values of $\pi_0$ and $\pi_1$. This is probably because the arcsine transformation is designed to stabilize power under the assumption that $\pi_0 = \pi_1$.

*Summary.* Under the assumption that a power-stabilising frontier holds, procedure 4, i.e. "conditionally modify margin" with a threshold $\epsilon = 1.25\%$ on the risk difference scale or $\epsilon =$ 1.25 on the risk ratio scale, is the best procedure. Power is higher than the "do not modify margin" procedure in almost all scenarios, and type I error is inflated only with the risk difference scale. We next explore two ways to control type I error in this case.

### 3.5. Controlling type I error rate

*(i) Smaller fixed α.* The simplest way of controlling type I error is to widen the confidence intervals using a smaller significance level $\alpha$ than the nominal 2.5% (for a one-sided test). We investigate this approach by repeating the base-case simulations for the risk difference, using different significance levels with procedure (4), the smallest threshold for margin modification.

Type I error is always below or around the nominal 2.5% level when using $\alpha = 1\%$ (Figure 4); this leads to a further loss in power of around 8-9% compared to the "do not modify margin" method. In general, conclusions depend on the relation between expected and observed control event risk:

- $\pi_0 < \pi_{e0}$: the "conditionally modify margin" procedure with $\alpha = 1\%$ is the only one with type I error within 2.5%;

- $\pi_0 = \pi_{e0}$: the original sample size calculation was correct, and hence the "do not modify margin" procedure performs well, while the "conditionally modify margin" procedure with smaller $\alpha$ loses ~10-15% power;

- $\pi_0 > \pi_{e0}$: the "do not modify margin" procedure quickly loses power, while all the "conditionally modify margin" procedures are quite stable and have correct type I error.

*(ii) Choose α given control risk.* Whilst one might simply recommend the "conditionally modify margin" procedure with $\alpha = 1\%$, this approach may be unnecessarily conservative for control event risks where larger $\alpha$ still leads to good type I error. Hence, another approach could be to choose $\alpha$ after observing the control event risk, using the largest $\alpha$ leading to acceptable type I error for that specific value of the control event risk. This can be estimated from simulations with the desired design parameters analogous to Figure 4. However, since $\alpha$ is chosen in a data-dependent way, this procedure is not guaranteed to preserve type I error. Nevertheless estimating the type I error from the previous simulations shows the inflation is at most modest (Figure 5), and hence this approach could be considered acceptable in practice, although it still leads to a 5-10% loss in power.

A simple way to prevent the additional loss of power is to design the trial using either the smaller fixed $\alpha$ with method (i) or $\alpha$ at $\pi_{e0} = \hat{\pi}_0$ with method (ii).

## 4. Discussion

We have addressed the challenge of designing a non-inferiority trial that preserves power and interpretability of results even when the expected control event risk is badly misjudged. Whilst statistically one could argue that sample size re-estimation based on interim analysis, updating the control group event risk and maintaining the original non-inferiority margin, solves this problem, in practice substantial increases in sample size are typically not acceptable to funders and may also be challenging for recruitment. Additionally, we argue that keeping the margin fixed may not be the optimal choice for the interpretation of results. Therefore alternative statistically principled methods are needed.

We have proposed three methods based on the definition of a non-inferiority frontier. Recently, Hanscom et al.[14] proposed using baseline or post-randomisation data to re-estimate the non-inferiority margin where this is based on preserving a fraction of the control group effect. Our methods are an alternative that can be pre-specified at the trial design stage.

## 4.1. Extensions

We have considered only binary outcomes, with risk differences and risk ratios as effect measures. Our approach could easily incorporate other effect measures, such as odds ratios or averted infection ratios[15], either to define an alternative non-inferiority frontier, or as the basis of a "conditionally modify margin" procedure assuming the power-stabilising frontier. Similar considerations could be extended to time-to-event outcomes. Again, a non-inferiority frontier could be chosen for absolute differences (e.g. Kaplan-Meier estimates of proportion after a certain time) or relative differences (e.g. hazard ratio).

Non-inferiority trials can have continuous outcomes, for example, the Early Treatment Diabetic Retinopathy Study score (number of letters a patient can read off a chart from a certain distance) in the CLARITY trial[16]. The investigators used an absolute non-inferiority margin of 5 letters, corresponding to a constant difference non-inferiority frontier. This is appropriate if the margin is independent of the control group mean. Otherwise, if the minimum acceptable number of letters depended on the control group mean, a relative difference, e.g. the ratio of the scores, might be used. However, an important difference compared to binary outcomes is that the sample size (and hence power) calculations for trials with continuous outcomes are independent of the expected control group mean when the variance is not associated with the mean. Hence, power is naturally preserved when assuming a fixed difference frontier.

Future work could investigate how to choose the modification threshold $\epsilon$ optimally when using the "conditionally modify margin" method.

## *4.2. Recommendations*

Given our results, researchers designing non-inferiority trials with a binary or time-to-event outcome should carefully consider the following:

1. The scale on which the non-inferiority comparison is made should be pre-specified in the trial protocol, as it substantially affects trial power (and hence sample size);

2. It is not obvious that the non-inferiority margin should be held fixed (on either risk difference or risk ratio scale) when $\hat{\pi}_0$ differs from the expected $\pi_{e0}$. Keeping it fixed could have implications in terms of power and interpretation, and these need to be considered carefully;

3. A trial design should explicitly pre-specify a "non-inferiority frontier", i.e. a curve indicating the tolerable non-inferiority margin for each value of the control event risk. This might be as simple as stating that the non-inferiority margin is fixed on the chosen scale;

4. One choice of non-inferiority frontier is based on the arcsine transformation. Although difficult to interpret per-se, this has the advantage of being the power-stabilising frontier for binomially distributed data;

5. One approach is to test on the arcsine scale and report results on the risk difference scale. However, this generally requires larger sample sizes. Testing on the arcsine scale and reporting on the risk-ratio scale is not recommended as it leads to reporting results against large margins or significance levels;

6. An alternative implementation is via our proposed "conditionally modify margin" procedure, which re-assesses the margin after observing the control event risk. The trial is still designed and analysed in the usual way, using either a risk difference or a risk ratio margin;

7. When using the "conditionally modify margin" procedure, an appropriate modification threshold can be selected through simulations as here. We will make available functions to perform such simulations in the R package `dani`.

8. If working on the risk difference scale, type I error rate should be controlled using simulations as here to find the appropriate nominal significance level. A conservative approach uses the largest level leading to a rate always below the nominal one, irrespective of the control event risk; otherwise, one can use simulation results to modify the significance level depending on the observed control event risk.

### *4.3. Conclusions*

Our proposed method of designing non-inferiority trials through defining a non-inferiority frontier and possibly modifying the non-inferiority margin accordingly after observing the control event risk substantially increases their resilience to inadvertent misjudgments of the control group event risk. The only disadvantage of this method is that, when working on the risk difference scale, some loss of power is expected, and hence sample size should be adjusted accordingly. Explicitly acknowledging before a trial starts that there could be differences between observed and expected control event risks forces researchers to focus in greater depth on the rationale underpinning their choice of non-inferiority margin, and the consequences to the trial if they get these assumptions wrong. We consider that researchers following our recommendations while designing non-inferiority trials with a binary primary outcome will improve the chance that the trial achieves its aims and will make it resilient to unexpected differences in the control event risk.

**Funding**

This work was supported by the Medical Research Council [MC_UU_12023/29].

**ORCID iD**

Matteo Quartagno https://orcid.org/0000-0003-4446-0730

389: 2193–2203.

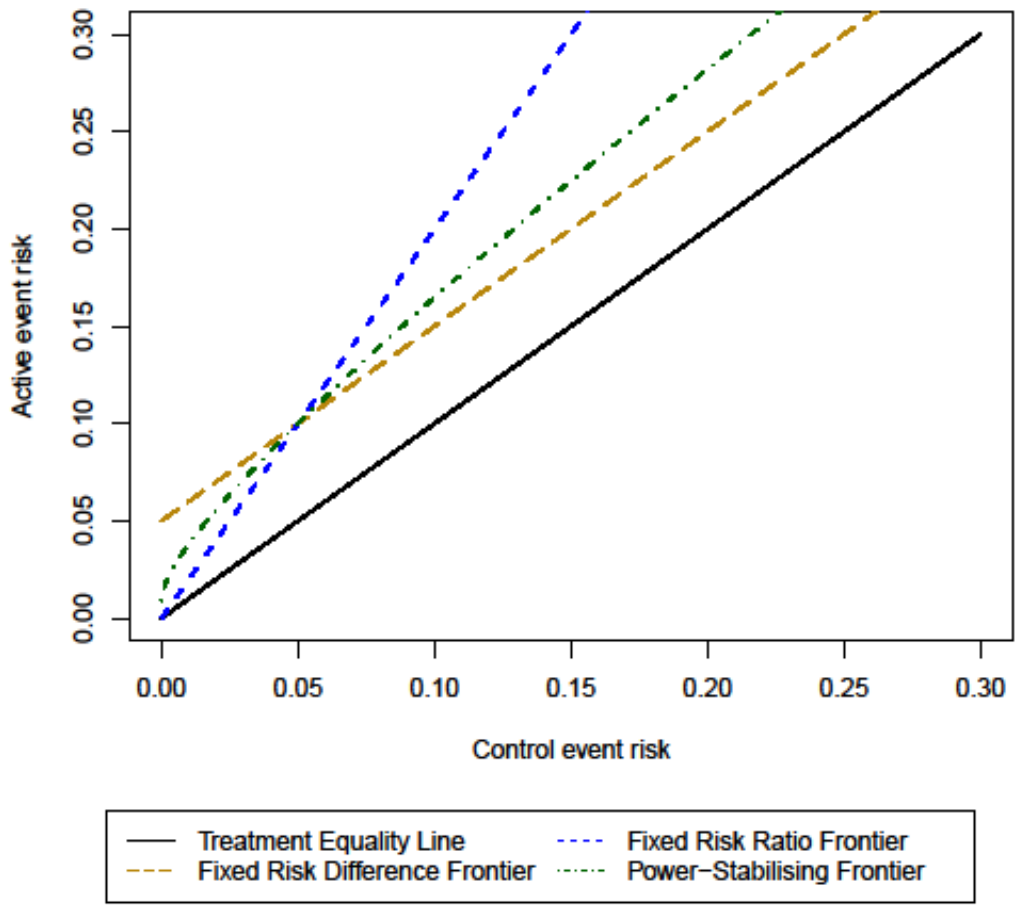

Figure 1: Non-inferiority frontiers: comparison of fixed risk ratio (2), fixed risk difference (5%) and power-stabilising frontiers. The black solid line corresponds to strict equivalence of the two treatments.

Table 1: Design parameters of the different simulation scenarios. $\pi_{e0}$ and $\pi_{e1}$ represent the expected control and active event risk, $\pi_{f1}$ the maximum tolerable active event risk and r the allocation ratio,

| Scenario | $\pi_{e0}$ | $\pi_{e1}$ | $\pi_{f1}$ | $r$ | Power |
|---|---|---|---|---|---|
| Base-Case | 5% | $= \pi_{e0}$ | 10% | 1 | 90% |
| Alternative 1 | **10%** | $= \pi_{e0}$ | **15%** | 1 | 90% |
| Alternative 2 | 5% | $= \dfrac{\pi_{e0}}{2}$ | 10% | 1 | 90% |
| Alternative 3 | 5% | $= \pi_{e0}$ | **7.5%** | 1 | 90% |
| Alternative 4 | 5% | $= \pi_{e0}$ | **15%** | 1 | 90% |
| Alternative 5 | 5% | $= \pi_{e0}$ | 10% | **0.5** | 90% |
| Alternative 6 | 5% | $= \pi_{e0}$ | 10% | **2** | 90% |
| Alternative 7 | 5% | $= \pi_{e0}$ | 10% | 1 | **80%** |

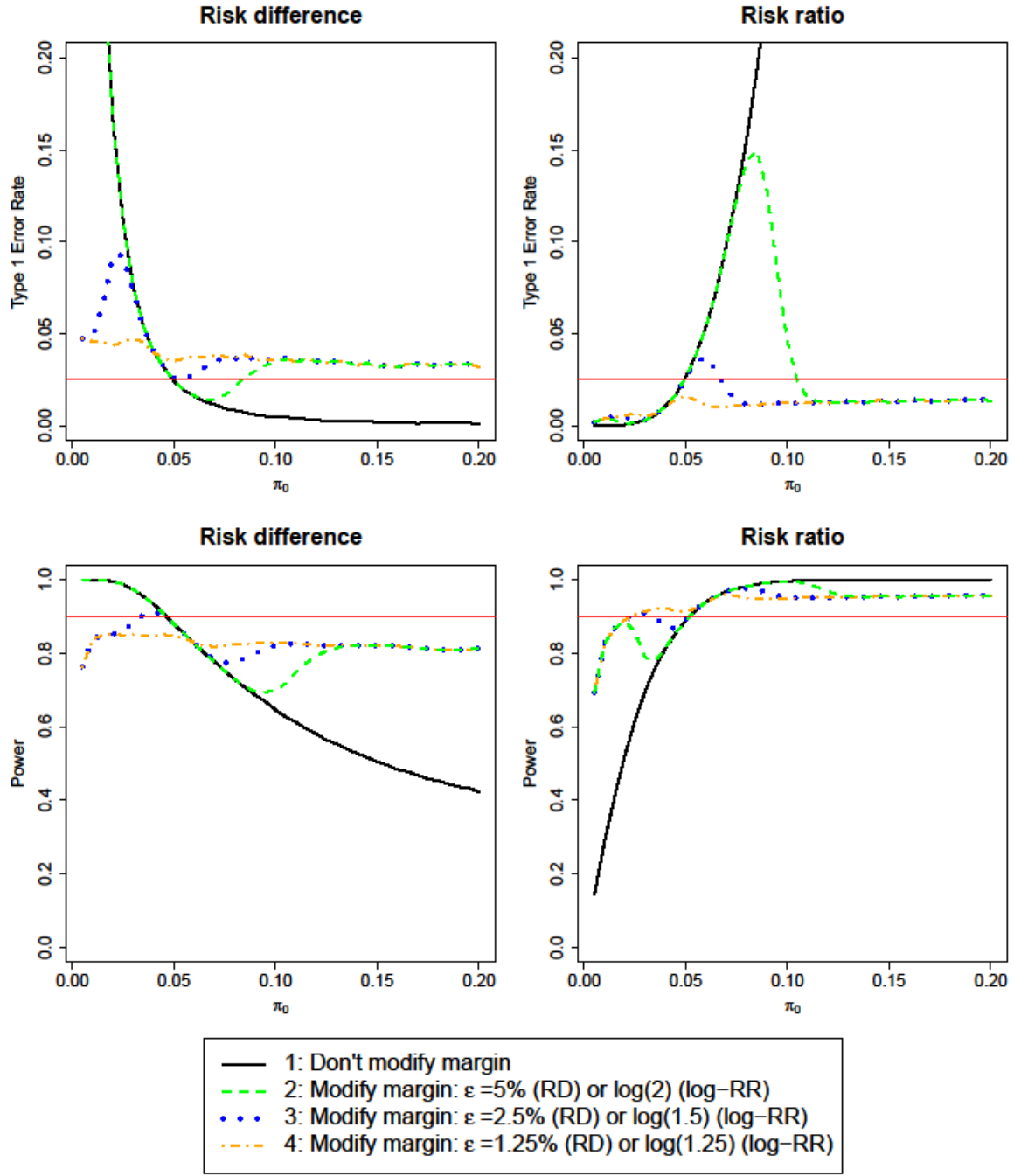

Figure 2: Type I error (top) and power (bottom) of "do not modify margin" and "modify margin" procedures, using the risk difference (left) or risk ratio (right) scale. Data are generated according to the base-case scenario of Table 1 for varying values of control event risk.

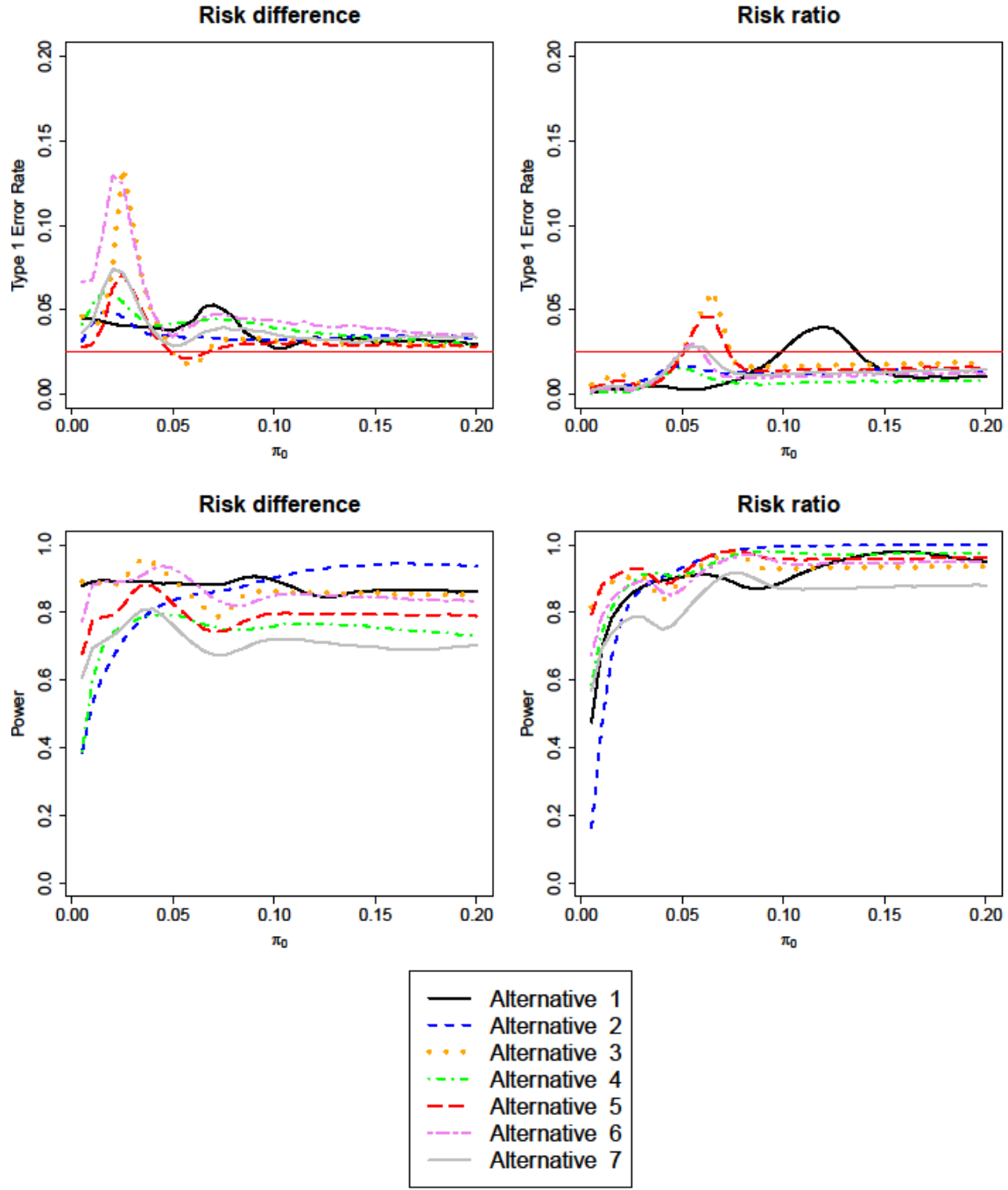

Figure 3: Type I error (top) and power (bottom) of the "conditionally modify margin" procedure, using the risk difference (left) or risk ratio (right) scale. Data are generated according to the alternative scenarios of Table 1 for varying values of control event risk.

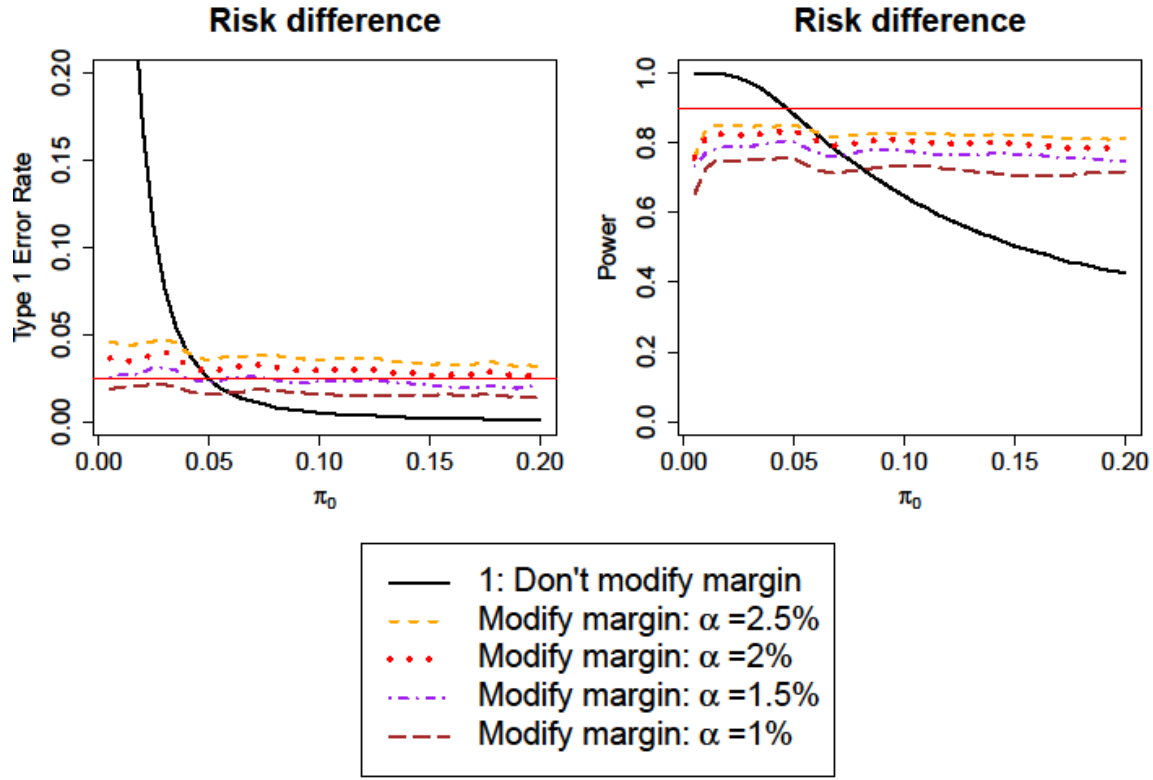

Figure 4: Power and type I error of procedure 4 ("Conditionally modify margin with small threshold"), with different significance levels. Only presenting the risk difference case, as type I error of the base-case scenario was below the nominal 2.5% level when working on the risk ratio scale.

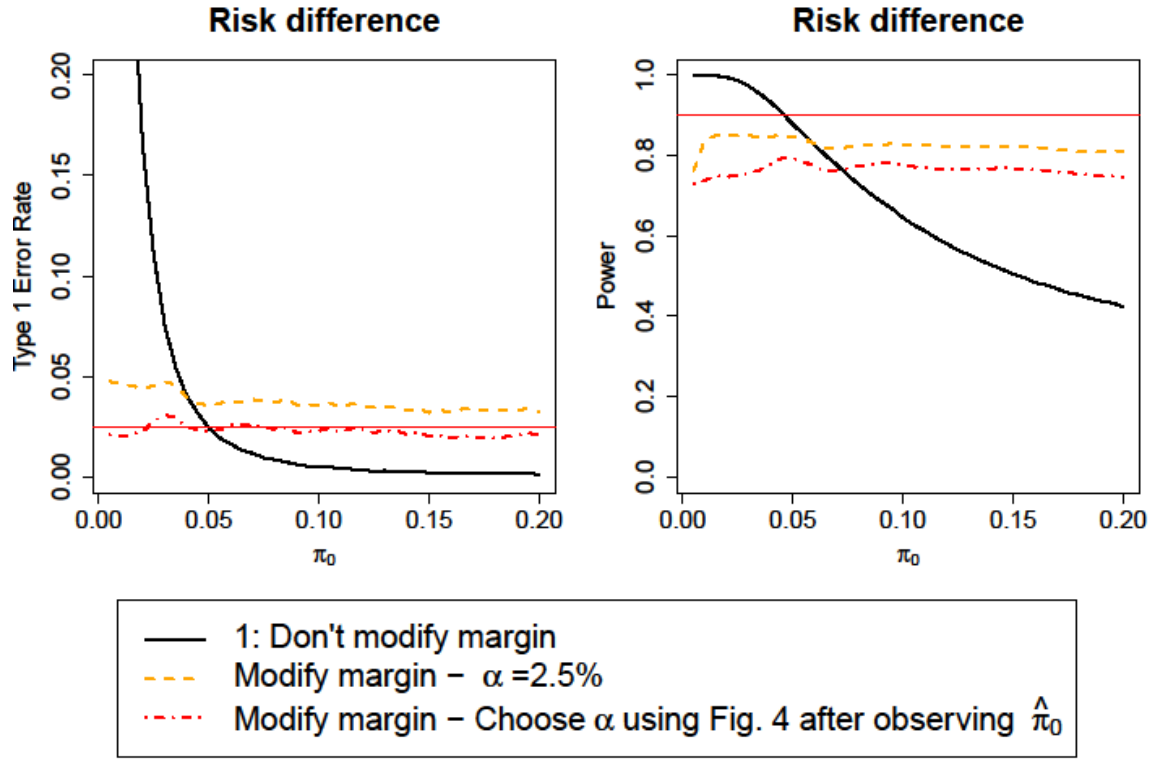

Figure 5: Power and type I error rate of procedure 4 ("Conditionally modify margin with smallest threshold"), either with standard significance level (one-sided α=2.5%) or choosing significance level using Figure 4 after observing control event risk $\hat{\pi}_0$ to achieve nominal type I error rate; specifically, in this example we use α=1% for $\hat{\pi}_0 < 4\%$ and α=1.5% otherwise.

## Additional material:

## Appendix A: Sample size calculation formulas

$n_0$ = sample size control arm

$n_1$ = sample size active arm

$\Phi^{-1}$ = standard normal quantile function

$\alpha$ = significance level

$\beta$ = type-II error

$\pi_{e0}$ = expected risk in the control arm

$\pi_{e1}$ = expected risk in the active arm

$r = \frac{n_1}{n_0}$ = allocation ratio

$\delta$ = Non-inferiority margin

**Risk difference:**

$$n_0 = (\Phi^{-1}(1-\alpha) + \Phi^{-1}(1-\beta))^2 \frac{((\pi_{e0} * (1-\pi_{e0}) + \pi_{e1} * (1-\pi_{e1})/r)}{(\pi_{e1} - \pi_{e0} - \delta)^2}$$

$$n_1 = r * n_0$$

**Risk ratio:**

$$n_0 = (\Phi^{-1}(1-\alpha) + \Phi^{-1}(1-\beta))^2 \frac{(((1-\pi_{e0})/(\pi_{e0}) + (1-\pi_{e1})/(r*\pi_{e1})))}{(log(\pi_{e1}/\pi_{e0}) - \delta)^2}$$

$$n_1 = r * n_0$$

**Arc-sine difference:**

$$n_0 = (\Phi^{-1}(1-\alpha) + \Phi^{-1}(1-\beta))^2 \frac{(\frac{1}{4} + \frac{1}{4r})}{(asin(\sqrt{\pi_{e1}}) - asin(\sqrt{\pi_{e0}}) - \delta)^2}$$

$$n_1 = r * n_0$$

# Appendix B: Illustrative design and analysis examples

In this appendix we provide examples of the design and analysis of hypothetical trials following the methods presented in this paper. We compare the results with those from the analysis of a trial assuming either a fixed risk difference or risk ratio frontier. For all the examples, design parameters are as in the base-case scenario of our simulation study, i.e. $\pi_{e0} = \pi_{e1} = 5\%$, $\pi_{f1} = 10\%$, power = 90%, one-sided $\alpha = 2.5\%$ and r = 1.

*Design*

As described in the main text, because of the different shapes of the non-inferiority frontier, with these same design parameters, the estimated total sample sizes for standard non-inferiority trials designed with a fixed risk difference, fixed risk ratio, and fixed arc-sine difference are 800, 1664 and 1136 respectively.

To incorporate resilience to unanticipated variation in the control event risk into a design using the fixed risk difference scale by conditionally modifying the risk difference margin with a threshold $\epsilon = 1.25\%$ requires either a data-dependent choice of one-sided alpha at the end of the trial or a more conservative upfront lowering of the one-sided alpha to 1%. As α=1.5% is the acceptable significance level when the expected and observed control event risk match, sample size is inflated to 903 using the first method (13% increase from the standard sample size calculation on the risk difference scale α=2.5%), while for the second method this goes up to 990 (23% increase). However, as the margin is wrongly modified ~20% of the times using the smallest threshold ε, actual power may be slightly lower than the nominal level. Future work will explore how to perform a more precise sample size calculation.

*Analysis*

For simplicity, we use the same dataset for all analysis methods; although some designs require different sample sizes as above, here we illustrate the methods using the sample size required to reach 90% power to prove non-inferiority on the arc-sine difference scale within $\text{asin}(\sqrt{\pi_{f1}}) - \text{asin}(\sqrt{\pi_{e0}}) = 0.096$, , i.e. 568 patients per arm, total 1136 patients.

We show how to analyse a trial with each of the following methods:

- Test and report on the arc-sine scale, as in Section 3.1;

- Test on the arc-sine scale, report on the risk difference scale changing the margin, as in Section 3.2(i);

- Test on the arc-sine scale, report on the risk difference scale changing significance level and modifying the margin, as in Section 3.2(ii);

- Test and report on the risk difference scale, as per a standard non-inferiority trial;

- Test on the risk difference scale, modifying the margin if $|\hat{\pi}_0 - \pi_{e0}| > 1.25\%$ and testing with $\alpha = 1\%$, as in Section 3.5(i).

- Test on the risk difference scale, modifying the margin if $|\hat{\pi}_0 - \pi_{e0}| > 1.25\%$ and choosing $\alpha$ using Figure (4), as in Section 3.5(ii).

- Test and report on the risk ratio scale, as per a standard non-inferiority trial on this scale;

- Test on the risk ratio scale, modifying the margin if $|\log(\hat{\pi}_0/\pi_{e0})| > \log(1.25)$ and testing with $\alpha = 2.5\%$, as in Section 3.4.

*(i) Example 1: actual control and intervention event risks are higher than anticipated*

First we consider the analysis of a trial where the observed event risks in the control and active arms are 5% higher than $\pi_{e0}$ and $\pi_{e1}$ respectively, i.e. $\hat{\pi}_0 = 10\%$, $\hat{\pi}_1 = 10\%$. In our hypothetical example trial, we observe 57 events in both the control ($\hat{\pi}_0 = 10\%$) and active arm ($\hat{\pi}_1 = 15\%$).

**Test and report on the arc-sine scale.** The estimated arc-sine difference and associated standard error are:

$$\widehat{AS} = \mathrm{asin}(\sqrt{\hat{\pi}_1}) - \mathrm{asin}(\sqrt{\hat{\pi}_0}) = 0.000 \qquad \hat{\sigma}_{AS} = \sqrt{\frac{1}{4n_0} + \frac{1}{4n_1}} = 0.030$$

Hence, the Z statistic is:

$$Z_{AS} = \frac{\widehat{AS} - \delta_{AS}}{\hat{\sigma}_{AS}} = -3.244$$

So, the p-value for the test of non-inferiority within margin 0.096 gives a p-value = $\Phi^{-1}(Z_{AS}) < 0.01$, providing evidence that the new treatment is non-inferior. The two-sided 95% confidence interval for the arc-sine difference is: [-0.058; 0.058].

**Test on the arc-sine scale, report on the risk difference scale changing the margin.** Since the test is performed on the arc-sine scale, and the trial has been designed on this scale, the Z statistic and p-value are the same as above. The only difference is that we report the results on the risk difference scale by calculating the non-inferiority margin leading to the same Z statistic. Following Section 3.2(i), we first estimate the risk difference and its associated standard error:

$$\widehat{RD} = \hat{\pi}_1 - \hat{\pi}_0 = 0.0\% \qquad \hat{\sigma}_{RD} = \sqrt{\frac{\hat{\pi}_0(1-\hat{\pi}_0)}{n_0} + \frac{\hat{\pi}_1(1-\hat{\pi}_1)}{n_1}} = 1.8\%$$

And then we find for which non-inferiority margin $\delta_{RD}$ these lead to the same Z statistic as $Z_{AS}$:

$$\delta_{RD} = \widehat{RD} - Z_{AS} * \hat{\sigma}_{RD} = 5.8\%$$

Hence, we report that we found the new treatment was non-inferior within the 5.8% risk difference margin, with p<0.01 and two-sided 95% confidence interval [-3.5%; 3.5%].

**Test on the arc-sine scale, report on the risk difference scale changing significance level.** Here, the estimated risk difference and associated standard error are the same as for the previous method. However, the non-inferiority margin is back-calculated from the power-stabilising frontier as:

$$\delta_{RD}^* = sin(asin(\sqrt{\hat{\pi}_0}) + asin(\sqrt{\pi_{f1}}) - asin(\sqrt{\pi_{e0}}))^2 - \hat{\pi}_0 = 6.5\%$$

The Z statistic with this larger non-inferiority margin is:

$$Z_{RD} = \frac{\widehat{RD} - \delta_{RD}^*}{\hat{\sigma}_{RD}} = --3.639$$

Hence, the ratio between $Z_{RD}$ and $Z_{AS}$ is equal to 1.12, and $z_{1-\alpha^*} = 1.11 z_{1-\alpha} = 2.20$. This is true for one-sided $\alpha^* = 1.4\%$.

In conclusion, we proved non-inferiority within the 6.5% risk difference margin at the 1.5% one-sided significance level (p<0.01), with 97.2% two-sided confidence interval [-3.9%; 3.9%].

**Test and report on risk difference scale.** This is a standard non-inferiority trial assuming a fixed risk difference frontier. Although an improved method would be preferable here to perform the test due to better efficiency[1], we retain the simple test using normal theory, for simplicity and comparability with the other methods. The estimated risk difference and associated standard error are the same as with the previous methods, i.e. $\widehat{RD} = 0.0\%$ and $\hat{\sigma}_{RD} = 1.8\%$. However, here we test at the pre-defined 5% non-inferiority margin:

$$Z_{AS} = \frac{\widehat{RD} - \delta_{RD}}{\hat{\sigma}_{RD}} = -2.778 \qquad p < 0.01$$

Finally, we report the 95% confidence interval for the risk difference, i.e. [-3.5%; 3.5%], and conclude that we provided evidence of non-inferiority at the 5% risk difference margin (p=0.50).

**Test on the risk difference scale with α=1%, modifying the margin.** Here, we start again from the same estimates of risk difference and associated standard error:

$$\widehat{RD} = 0.0\% \quad \hat{\sigma}_{RD} = 1.8\%$$

In this example, since $|\hat{\pi}_0 - \pi_{e0}| = 5\% > 1.25\%$, we modify the margin from the initially intended 5% to the one back calculated from the arc-sine frontier given $\hat{\pi}_0$, i.e. 6.5%. Hence, the Z statistic and p-value are:

$$Z_{RD} = \frac{\widehat{RD} - \delta^*_{RD}}{\hat{\sigma}_{RD}} = -3.639 \qquad p < 0.01$$

In conclusion, we prove non-inferiority at one-sided 2.5% significance level within the 6.5% margin (p<0.01), and the two-sided 95% confidence interval for the risk difference is [-4.2%; 4.2%], larger than with the fixed risk difference example as we have used α=1% in order to control type 1 error.

**Test on the risk difference scale, modifying the margin and choosing α using Figure 4.** This method is similar to the previous one, but this time the significance level for testing is chosen from Figure 4, given that $\hat{\pi}_0 = 10\%$. In this specific case, this leads to using a one-sided α=1.5%, and hence the confidence interval is now [-3.9%; 3.9%].

**Test and report on the risk ratio scale.** The estimated log-risk ratio and associated confidence interval are:

$$\widehat{RR} = \log(\hat{\pi}_1) - \log(\hat{\pi}_0) = 0.00 \quad \hat{\sigma}_{RR} = \sqrt{\frac{(1-\hat{\pi}_0)}{n_0 \hat{\pi}_0} + \frac{(1-\hat{\pi}_1)}{n_1 \hat{\pi}_1}} = 0.18$$

The non-inferiority margin on the log-risk ratio scale is $\delta_{RR} = \log\left(\frac{0.1}{0.05}\right) = 0.69$, leading to the following Z statistic for the test on the log-risk ratio scale:

$$Z_{RR} = \frac{\widehat{RR} - \delta_{RR}}{\hat{\sigma}_{RR}} = -4.29 \qquad p < 0.01$$

There is therefore strong evidence at the one-sided 2.5% significance level that the new treatment is non-inferior to the control within a log-risk ratio margin of 0.69, i.e. that the relative risk is less than

2. Note that since we are working with a fixed risk ratio scale, and $\hat{\pi}_0 = 10\%$, we would now be happy to tolerate up to $\hat{\pi}_1 = 20\%$. In this case the 95% confidence interval should be reported on the risk ratio scale: [0.71, 1.42].

**Test and report on the risk ratio scale, modifying the margin.** Since $|\log(\hat{\pi}_0/\pi_{e0})| = \log(2) > \log(1.25)$, the margin is modified according to the power-stabilising frontier:

$$\delta_{RR}^* = \log(\sin(asin(\sqrt{\hat{\pi}_0}) + asin(\sqrt{\pi_{f1}}) - asin(\sqrt{\pi_{e0}}))^2) - \log(\hat{\pi}_0) = 0.50$$

The estimated log-risk ratio and standard error are as before:

$$\widehat{RR} = 0.00 \quad \hat{\sigma}_{RR} = 0.18$$

Differently from the risk difference case, in this situation there is no need to adjust for type-1 error rate, and hence we can keep one-sided α=2.5% as significance level. The Z statistic and p-value are:

$$Z_{RR} = \frac{\widehat{RR} - \delta_{RR}^*}{\hat{\sigma}_{RR}} = -3.11 \qquad p < 0.01$$

The 95% confidence interval on the risk ratio scale is [0.71; 1.42], and the p-value for the non-inferiority test within the $e^{0.5} = 1.65$ margin is < 0.01.

*(ii) Other examples.* Table (a) below shows the results of the analysis performed on the previous example and on two additional examples. These are all based on the same design parameters, i.e. those for the base-case scenario of our simulation study, leading to a sample size of 568 patients per arm. However, in the second example $\hat{\pi}_1 = 15\%$, so that the observed active and control event risk differ by 5%, while in the third example $\hat{\pi}_0 = 6\%$, $\hat{\pi}_1 = 10\%$, so that the active event risk is close to its expected value at the design stage, and hence we do not modify the margin when using methods from Section 3.4 and 3.5.

| Method | CI Level | CI | Margin | 1-sided α level | p |
|---|---|---|---|---|---|
| Example 1: $\hat{\pi}_0 = 10\%, \hat{\pi}_1 = 10\%$ | | | | | |
| Test&Report on Arc-sine scale | 95% | [-0.058; 0.058] | 0.096 | 2.5% | <0.01 |
| Test arc-sine, report RD (change margin) | 95% | [-3.5%; 3.5%] | 5.7% | 2.5% | <0.01 |
| Test arc-sine, report RD (change α) | 97.2% | [-3.9%; 3.9%] | 6.5% | 1.4% | <0.01 |
| Test RD | 95% | [-3.5%; 3.5%] | 5% | 2.5% | <0.01 |
| Test RD, Modify Margin, α=1% | 95% | [-4.2%; 4.2%] | 6.5% | 1% | <0.01 |
| Test RD, Modify Margin, choose α | 95% | [-3.9%; 3.9%] | 6.5% | 1.5% | <0.01 |
| Test RR | 95% | [0.71, 1.42] | 2 | 2.5% | <0.01 |
| Test RR, Modify Margin | 95% | [0.71, 1.42] | 1.65 | 2.5% | <0.01 |
| Example 2: $\hat{\pi}_0 = 10\%, \hat{\pi}_1 = 15\%$ | | | | | |
| Test&Report on Arc-sine scale | 95% | [0.018; 0.134] | 0.096 | 2.5% | 0.25 |
| Test arc-sine, report RD (change margin) | 95% | [1.2%; 8.8%] | 6.3% | 2.5% | 0.25 |
| Test arc-sine, report RD (change α) | 97.0% | [0.8%; 9.2%] | 6.5% | 1.5% | 0.25 |
| Test RD | 95% | [1.2%; 8.8%] | 5% | 2.5% | 0.50 |
| Test RD, Modify Margin, α=1% | 95% | [0.3%; 9.7%] | 6.5% | 1% | 0.25 |
| Test RD, Modify Margin, choose α | 95% | [0.8%; 9.2%] | 6.5% | 1.5% | 0.25 |
| Test RR | 95% | [1.10, 2.06] | 2 | 2.5% | 0.04 |
| Test RR, Modify Margin | 95% | [1.10; 2.06] | 1.65 | 2.5% | 0.29 |
| Example 3: $\hat{\pi}_0 = 6\%, \hat{\pi}_1 = 10\%$ | | | | | |
| Test&Report on Arc-sine scale | 95% | [0.016; 0.132] | 0.096 | 2.5% | 0.23 |
| Test arc-sine, report RD (change margin) | 95% | [0.9%; 7.2%] | 5.2% | 2.5% | 0.23 |
| Test arc-sine, report RD (change α) | 97.4% | [0.4%; 7.6%] | 5.4% | 1.3% | 0.23 |
| Test RD | 95% | [0.9%; 7.2%] | 5% | 2.5% | 0.27 |
| Test RD, Modify Margin, α=1% | 95% | [0.2%; 7.7%] | 5% | 1% | 0.27 |
| Test RD, Modify Margin, choose α | 95% | [0.5%; 7.5%] | 5% | 1.5% | 0.27 |
| Test RR | 95% | [1.11, 2.51] | 2 | 2.5% | 0.20 |
| Test RR, Modify Margin | 95% | [1.11; 2.51] | 2 | 2.5% | 0.20 |

**Table (a): Results of analysis of hypothetical trials using methods presented in this paper and standard non-inferiority designs on either the risk difference or risk ratio scales. For each method and example, we provide confidence level and corresponding interval, Non-inferiority margin, 1-sided significance level for testing and p-value. When using the "Modify Margin" methods, the 1-sided significance level used for testing can be modified without this affecting confidence level; this is because a different α is used only to maintain type-1 error below the nominal 2.5%, and hence the relevant confidence level is still two-sided 95%.**

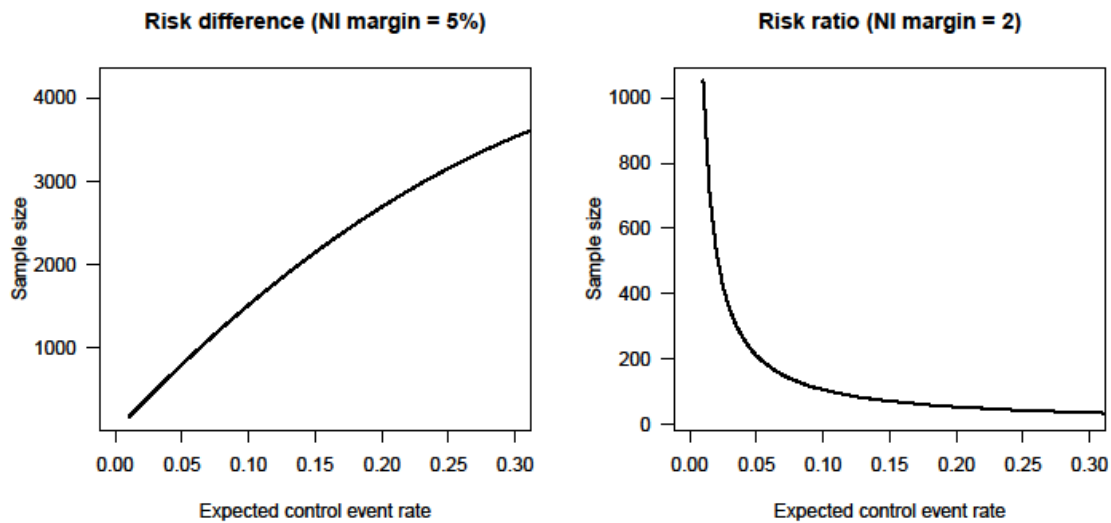

**Figure (a):** total sample size (2 groups) to achieve 90% power for varying control event risks using non-inferiority margins defined on the risk difference (left panel) and risk ratio (right) scales (two-sided alpha=0.05).

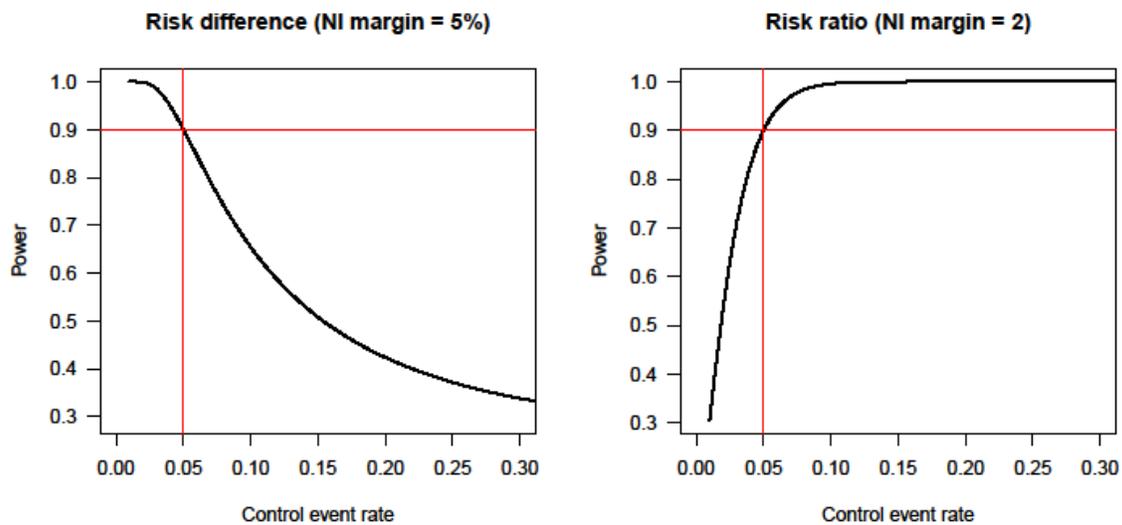

**Figure (b):** power for given sample size for varying control event risks using non-inferiority margins defined on the risk difference (N=400, left panel) and risk ratio (N=832, right panel) scales (two-sided alpha=0.05)

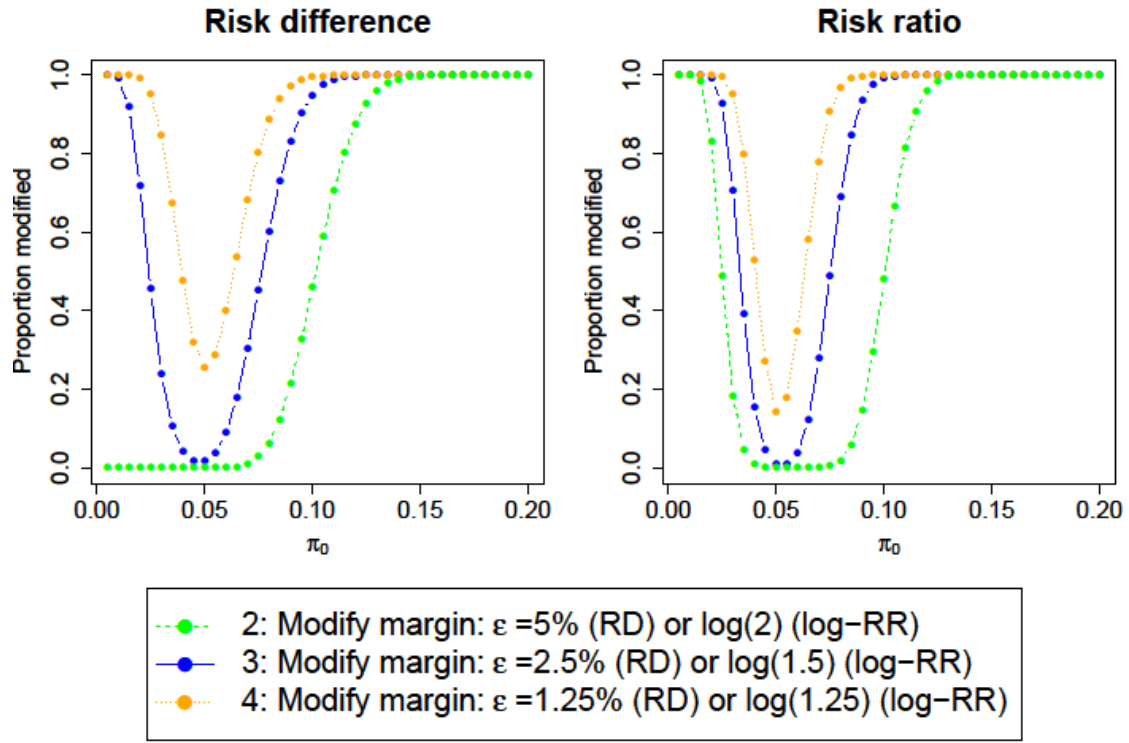

Figure (c): Proportion of margins modified using the three different "Conditionally modify margin" procedures. Data are generated according to the base-case scenario of Table 1 for testing type I error rate.

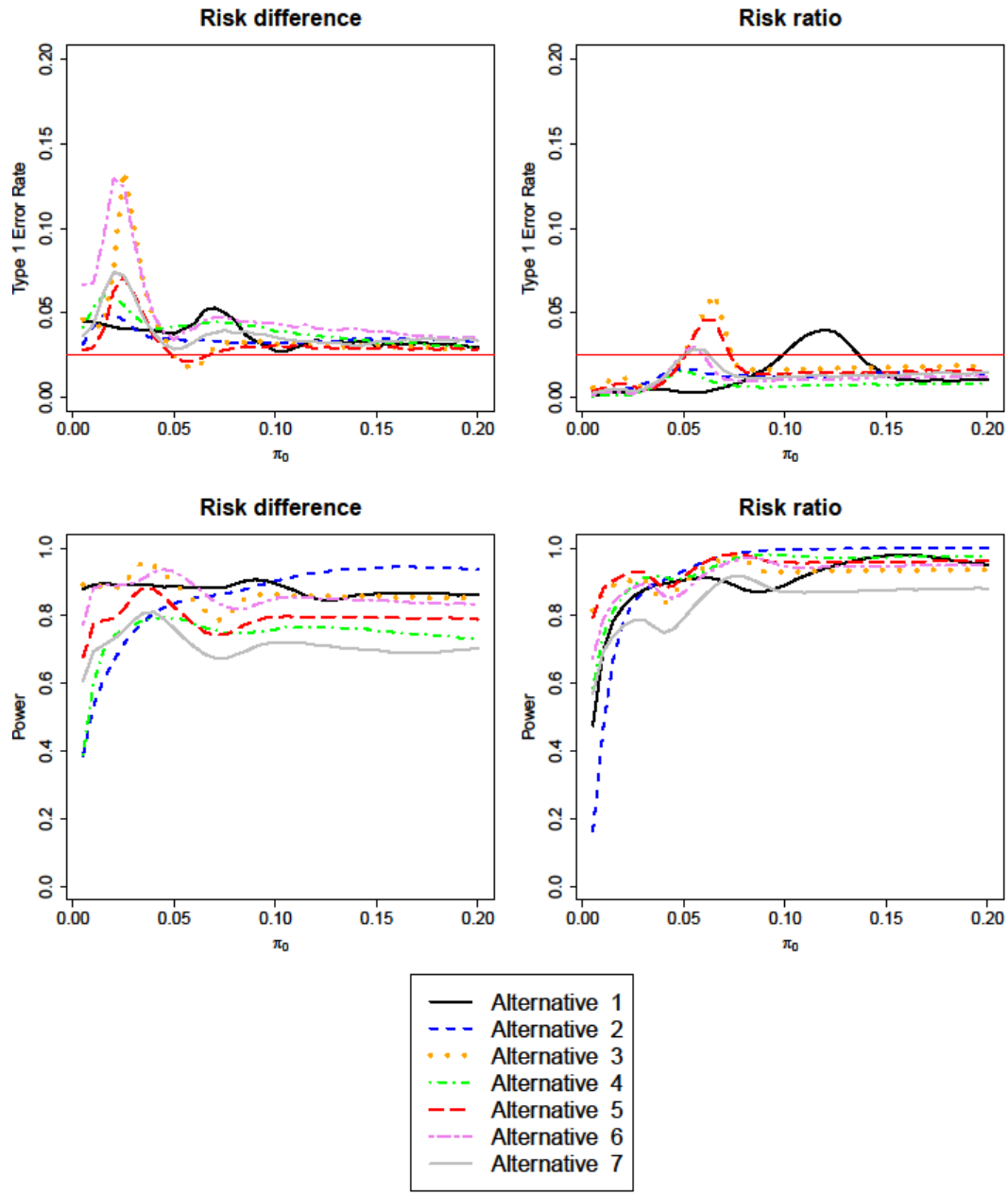

Figure (d): Type I error (top) and power (bottom) of procedure 3 ("Conditionally modify margin with medium threshold"), using the risk difference (left) or risk ratio (right) scale. Data are generated according to the alternative scenarios of Table 1 for varying values of control event risk.

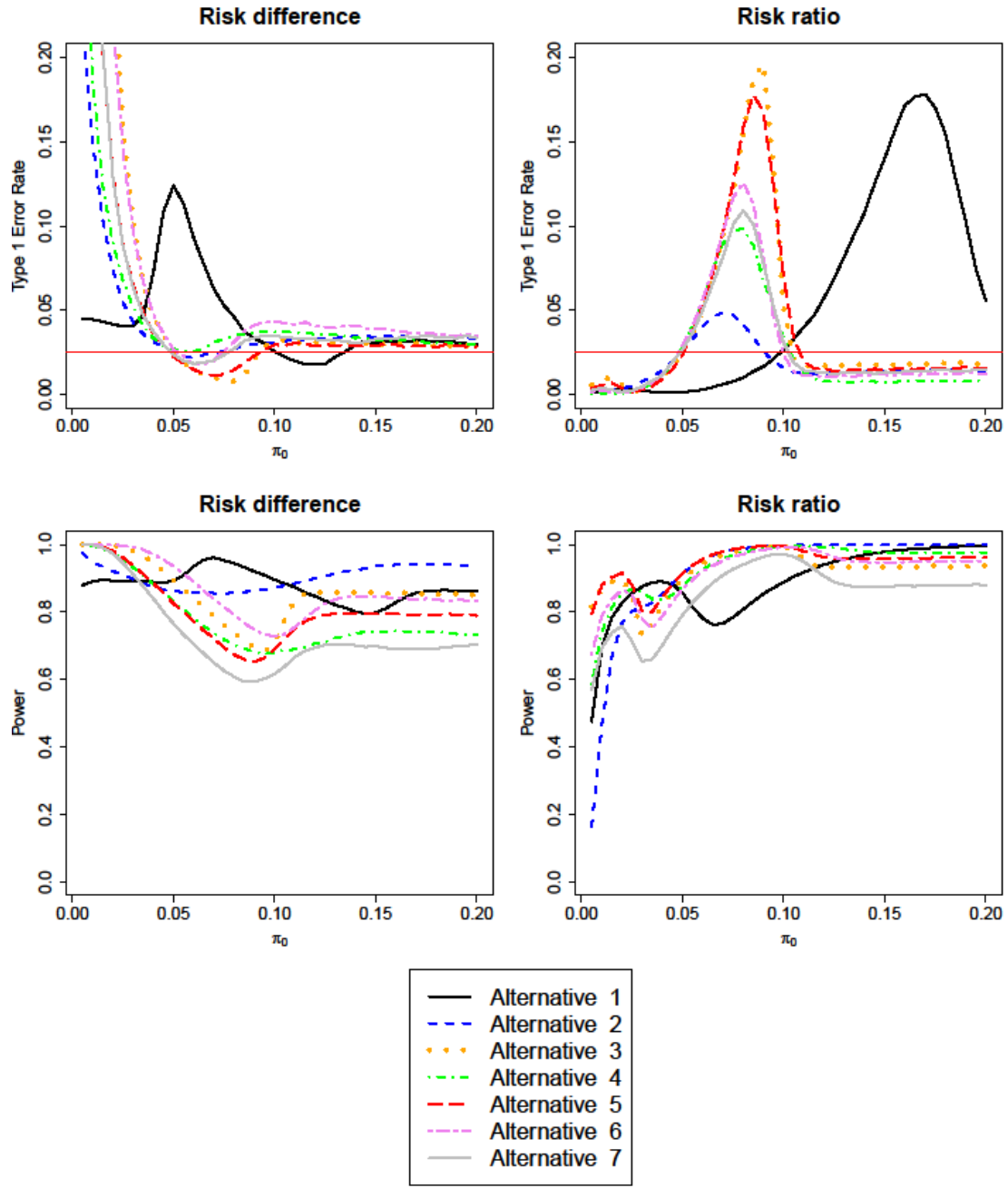

**Figure (e):** Type I error (top) and power (bottom) of procedure 2 ("Conditionally modify margin with largest threshold"), using the risk difference (left) or risk ratio (right) scale. Data are generated according to the alternative scenarios of Table 1 for varying values of control event risk.